# Coexistence of superconductivity and ferromagnetism in the graphite-sulfur system

S. Moehlecke[a], Pei-Chun Ho, and M. B. Maple

Department of Physics and Institute for Pure and Applied Physical Sciences,

University of California, San Diego, La Jolla, CA 92093-0360, U.S.A.

## Abstract

Superconducting characteristics such as the Meissner-Ochsenfeld state, screening supercurrents and hysteresis loops of type-II superconductors were observed from the temperature and magnetic field dependences of the magnetic moment, m(T, H), in graphite powders reacted with sulfur for temperatures below 9.0 K. The temperature dependence of the lower critical field $H_{c1}(T)$ was determined and the zero-temperature penetration depth, $\lambda(0)$, was estimated ($\lambda(0)$ = 2270 Å). The superconductivity was observed to be highly anisotropic and to coexist with a ferromagnetic state that has a Curie temperature well above room temperature. A continuous transition from the superconducting state to the ferromagnetic state could be achieved by simply increasing the applied magnetic field.

---

[a] On leave from Instituto de Física 'Gleb Wataghin', Universidade Estadual de Campinas, Unicamp 13083-970, Campinas, São Paulo, Brasil.

# 1. Introduction

Materials based on pure carbon and its allotropic forms have recently attracted a great deal of attention from the scientific community not only because of their unexpected properties, like ferromagnetism and superconductivity, but also due to the magnitude of these effects. Superconductivity (Schön et al. 2001) at temperatures as high as 117 K was obtained in electric-field hole doped $C_{60}$ fullerenes and ferromagnetism (Makarova et al. 2001) with large saturation magnetization and hysteresis loops was observed in polymerized $C_{60}$ which allow the samples to be attracted by a permanent magnet at room temperature. These spectacular properties that may not be limited to these values together with their potential applications have inspired a broad search in different carbon materials (pure and doped) for other phenomena. Highly oriented pyrolitic graphite (HOPG) is a good candidate and experimental results suggesting the occurrence of superconductivity at high temperatures were recently published (Kopelevich et al. 1999, 2000). However, the superconducting state generated in these samples was metastable and/or sometimes difficult to reproduce. Very recently, more stable superconducting results were reported in HOPG reacted with sulfur (Ricardo da Silva et al. 2001, Hai-Peng et al. 2001). Here we not only confirm these latter results and present additional superconducting characteristic parameters of this new graphite-sulfur system but also show that the superconductivity is anisotropic and intimately coexists with a ferromagnetic state that has a Curie temperature well above room temperature.

# 2. Experimental Details

The graphite-sulfur samples were prepared using graphite rods from Carbon of America Ultra® Carbon, AGKSP grade, Ultra 'F' purity (99.9995%) (Alfa-Aesar®, #

40766) and sulfur chunks from American Smelting and Refining Co. - ASARCO, spectrographically pure (99.999+%). A pressed pellet ($\phi$ = 6 mm) of graphite was prepared by pressing graphite powder at ~ 7 000 lb., the graphite powder was produced by cutting and grinding the graphite rod on the edge and side area of a new and clean circular diamond saw blade. The graphite pellet was encapsulated with sulfur chunks (mass ratio ~ 1:1) in a quartz tube under 1/2 atmosphere of argon and heat treated in a tube furnace at 400 °C for one hour and then slowly cooled (4 °C/h) to room temperature. X-ray diffraction measurements ($\theta$-2$\theta$ geometry and rocking curves) of the reacted sample yielded a spectrum with only the superposition of the (00$\ell$) diffraction peaks of graphite with the orthorhombic peaks of sulfur with no extra peak due to a compound, second phase or impurity. Also, the c-axis lattice parameter (c = 6.72 Å) of the sample is equal to the pristine graphite powder pellet, which suggests that no sulfur intercalation occurred during the reaction. The diffraction pattern also shows a strong (00$\ell$) preferred orientation which was confirmed by rocking curve scans that give a $\Delta\theta$ = 6° (FWHM) for the (002) peak, due to the highly anisotropic shape of the graphite grains. Optical microscopy analyses of the transverse cross section of the pellet reveal that the sample expanded significantly with the reaction (the pellet thickness more than doubled) forming several cracks and voids that are filled with pure sulfur. These sulfur inclusions are the main reason for the large weight increase observed after the reaction that nearly double the sample mass. Two pieces (~ 5 x 2.5 x 1.7 mm$^3$) of the sample pellet were cut and used for the magnetic and transport properties measurements and the above described analyses. The magnetic and electrical transport properties of the sample were measured using an MPMS5 SQUID magnetometer (Quantum Design®) and a PPMS (Quantum Design®), respectively.

Several other graphite-sulfur samples were prepared using the graphite rods and also HOPG from different sources. About a dozen of these samples exhibit superconductivity with different $T_c$ values; the sample with the largest superconducting volume fraction is described in this report. Even though some degree of reproducibility has been achieved in the sample preparation, not all of the conditions and parameters have been identified or optimized. However, it seams to be important for the occurrence of superconductivity that the graphite be activated by powdering (Harker et al. 1971) before the sulfur reaction since no bulk graphite could be turned superconducting. On the other hand, no correlations between the superconducting parameters and the initial graphite/sulfur mass ratio or the sample final mass were observed. Some samples exhibit no detectable change in mass after the reaction and even so become superconducting. The superconductivity and the ferromagnetism of the samples were observed to be stable from one week to ~ 6 months, and they can decrease as well as increase in magnitude abruptly with aging. The samples were usually stored in an argon atmosphere.

## 3. Results and discussion

Figure 1(a) shows the temperature dependence of the magnetic moment, m(T), normalized to the sample weight for two different magnetic fields applied perpendicular to the largest surface of the sample (H // c). Zero-field-cooled (ZFC) measurements, $m_{ZFC}(T)$, were made on heating after the sample was cooled in zero applied field to low temperatures and the desired magnetic field was applied. The field-cooled on cooling (FCC) measurements, $m_{FCC}(T)$, were made as a function of decreasing temperature for the same applied field. From figure 1(a), it can be seen that below 9 K both $|m_{ZFC}(T)|$ and $|m_{FCC}(T)|$ show a increase of the diamagnetic signal at the same temperature and for lower temperatures $|m_{ZFC}(T)| > |m_{FCC}(T)|$. This behavior is typical of a superconductor with a superconducting transition temperature $T_c = 9$ K, and the observed enhancement of the diamagnetic signal below $T_c(H)$ come from screening supercurrents generated in the sample during the ZFC measurements and the magnetic flux expulsion, the Meissner-Ochsenfeld effect, during the FCC measurements. The superconducting transition width $\Delta T_c$ (90% - 10%) is ~ 1.5 K for $H = 10$ Oe but it very quickly broadens with increasing magnetic field as also shown in figure 1(a) for $H = 70$ Oe. The two sample pieces cut from the pellet display the same value of $T_c$ and the superconducting characteristics described above. This suggests that the superconductivity of the graphite-sulfur sample is rather homogeneous, although large sulfur inclusions were also present. Note also from figure 1(a) that, in the normal state, the magnetic signal is already diamagnetic, due to the strong orbital diamagnetism of graphite (Kelly 1981, Heremans et al. 1994). Figure 1(b) shows the temperature dependence of the ZFC normalized magnetic moment $m(T)/|m(10 \text{ K})|$ measured for several different magnetic fields (H // c). The superconducting transition temperature $T_c(H)$, as indicated by arrows in figure 1b, decreases with increasing applied magnetic field, characteristic of superconducting materials.

Shown in figure 2 are magnetic moment hysteresis loops m(H) measured with the ZFC procedure (H // c) for T = 2 K, 5 K and 6 K, after subtraction of the diamagnetic background signal ($m_o = \chi H$, where $\chi(2 \text{ K}) = -7.14 \times 10^{-6}$ emu g$^{-1}$ Oe$^{-1}$; $\chi(5 \text{ K}) = -7.02 \times 10^{-6}$ emu g$^{-1}$Oe$^{-1}$ and $\chi(6 \text{ K}) = -6.96 \times 10^{-6}$ emu g$^{-1}$ Oe$^{-1}$). These hysteresis loops are typical of type II superconducting materials. Note the almost horizontal or constant behavior of the magnetic moment with field m(H) for the descending branches of these loops which resemble those superconductors where the vortex dynamics are dominated by surface pinning and no potential barrier exists for the exit of vortices (Bean and Livingston 1964). Also note that these loops show the occurrence of a spontaneous magnetic moment at zero applied magnetic field; this zero-offset has been observed before (Kopelevich et al. 2000) in graphite samples. The inset of figure 2 shows an example of the direct m(H) data measured for T = 2 K (H // c) where one can clearly observe the superposition of the superconducting hysteric loop over a linear diamagnetic background. We include in this figure 2 some data (★ for T = 5 K and ■ for T = 6 K) taken from the ZFC and FCC magnetic moment m(T) curves in different magnetic fields (some of the m(T) curves are shown in figures 1 (a) and (b)) that agree very well with the measured m(H) hysteresis loops, after adding their respective diamagnetic backgrounds and spontaneous magnetic moment. This good agreement illustrates the self-consistency of these measurements and the simple superposition nature of the diamagnetic and superconducting characteristics.

Presented in figure 3(a) are the initial low field portions of the m(H) hysteresis loops for different temperatures (ZFC, H // c), after subtraction of the diamagnetic background signal ($m_o = \chi$ H, where $\chi$ varies from -7.14 x $10^{-6}$ emu $g^{-1}$ $Oe^{-1}$ at 2 K to -6.5 x $10^{-6}$ emu $g^{-1}$ $Oe^{-1}$ at 7.5 K). All of the curves clearly show the common linear dependence of the magnetic moment on field associated with the Meissner effect. From these data the value of the lower critical field, $H_{c1}(T)$, can be estimated as the point of departure from linearity. The common Meissner line shown as a straight line in figure 3(a) was obtained from a linear fit of the data between 4 Oe and 50 Oe at 2 K. The temperature dependence of $H_{c1}(T)$ is shown in figure 3(b) where the continuous line is given by the relation

$$H_{c1}(T) = H_{c1}(0) [1 - t^{5/2}]^2, \qquad (1)$$

where $H_{c1}(0) = 64$ Oe and $t = T/T_c$ is the reduced temperature. Note that the $H_{c1}(T)$ curve has positive curvature near $T_c$, a characteristic of layered superconducting materials (Ruggiero et al. 1980). According to our measured hysteresis loops (see figure 2), these graphite-sulfur samples are type II superconductors with a relatively large Ginzburg-Landau parameter, $\kappa$. This allows us (Tinkham 1996) to consider $1/2(\ln \kappa) \approx 1$ and make a rough estimation of the penetration depth, $\lambda$, using the relation $H_{c1} \approx \phi_o/2\pi\lambda^2$. The calculated penetration depth as a function of temperature, $\lambda(T)$, is shown in the inset of figure 3(b); the continuous line is given by the equation $\lambda(T) = \lambda(0) [1 - t^{5/2}]^{-1}$, where $\lambda(0) \approx 2270$ Å for H // c. This value of $\lambda(0)$ is within the range of values of the penetration depth of the layered high-$T_c$ Bi(2212) superconductor in the optimally doped (2500 Å) and overdoped (1800 Å) states (Li et al. 1996). In figure 3(b) is also shown the field dependence of the onset of superconductivity, $T_c(H)$, determined from the m(T) curves (see figure 1). These values usually define the upper critical field boundary between the superconducting and the normal states and present a linear $(1 - t)$ dependence near $T_c$. However, our results are best described by the power law

$$H_b(T) = H^*(1 - t)^{3/2} \qquad (2)$$

with $H^* = 2.1$ T, as shown by the dashed line in figure 3(b). This kind of behavior suggests the existence of a breakdown field $H_b(T)$ that destroys the superconductivity induced by the proximity effect (Deutscher and de Gennes 1969, Deutscher and Kapitulnik 1990, Fauchere and Blatter 1997), and is the same behavior found previously in a graphite-sulfur sample with a higher $T_c$ (= 35 K) (Ricardo da Silva et al. 2001).

The results presented so far, for measurements made with the magnetic field applied perpendicular to the graphite planes (H // c), clearly show that the sample is in the superconducting state for T < 9.0 K. From the FCC m(T) measurements and from the initial slope of the hysteresis loops, we estimate a lower limit for the superconducting volume fraction of ~ 0.02%. Our transport measurements show that the sample has a room temperature resistivity $\rho(300K) = 0.3$ Ω-cm, a value much larger than that of the initial bulk graphite (0.7 mΩ-cm). The temperature dependence of the resistivity, $\rho(T)$, shows a semiconductor-like behavior and no anomaly was observed around 9 K.

The measurements with the applied magnetic field parallel to the graphite planes (H // a) present a completely different magnetic response, which is displayed in figure 4. Figure 4(a) shows the ZFC and FCC temperature dependencies of the magnetic moment m(T) for various applied fields, as indicated in the figure. No sign of a superconducting transition could be detected within this range of temperatures, even down to 2.0 K (not shown in figure 4(a)). Note that the scales in figure 4(a) are almost the same as in figure 1(a), and also that the diamagnetic signal is smaller in this field configuration (H // a) than when H // c. Shown in the upper inset of figure 4(b) is the direct hysteresis loop m(H) measured at 2 K (ZFC, H // a) which consist of a ferromagnetic-like hysteresis loop, clearly different from the loop shown in the inset of figure 2, that is superimposed on top of a linear diamagnetic background. Figure 4(b) shows the ZFC magnetic moment hysteresis loops m(H) for T = 2 K and 9 K after subtraction of the linear diamagnetic background ($m_o = \chi H$, where $\chi(2 K) = \chi(9 K) = -2.25 \times 10^{-6}$ emu g$^{-1}$ Oe$^{-1}$). Shown in the lower inset of figure 4(b) is the hysteresis loop measured at 5 K (H // a) up to higher fields where a trend towards saturation of the magnetic moment, $m_s$, can be seen ($\chi$(5 K) = -2.25 x $10^{-6}$ emu $g^{-1}$ $Oe^{-1}$). These hysteresis loops are typical of weak ferromagnetic materials and were observed in measurements up to room temperature. At the same time, no noticeable change or anomaly was observed in the hysteresis loops around 9 K. Figure 5(a) shows the temperature dependence (5 K - 350 K) of the remnant magnetic moment, $m_R$(T), after a field of 1000 Oe was applied at 15 K (H // a) and the sample was field cooled (FC) until 5 K, the applied field was removed and the remnant moment measured upon increasing temperature. From this figure it can be seen that the ferromagnetic state is still present at 350 K; extrapolation of the apparent linear behavior of $m_R$(T) between 5 K and 350 K to $m_R$ = 0 yield a Curie temperature of ~ 750 K (as a upper limit). Shown in figure 5(b) is the remnant moment from 5 K to 15 K for H // c and three FC values (500 Oe, 1000 Oe and 10 000 Oe), as indicated next to each curve in the figure. All three curves show a linear decrease of $m_R$(T) with a clear change in slope at ~ 9 K (the straight lines through the data of the figure emphasize this feature). This type of behavior of the $m_R$(T) curves support the existence of a superconducting state for T < 9 K and H // c. In general, a superconductor has a remnant moment temperature dependence, $m_R$(T), similar to the diamagnetic moment temperature dependence m(T), except the signs are opposite. From figure 1, it can be seen that the temperature dependence of the diamagnetic moment is almost linear for H > 30 Oe in this temperature interval, in agreement with the above observation. Also shown in figure 5(b) are the same initial remnant moment $m_R$(T) data of figure 5(a) from 5 K to 15 K (H // a). The temperature dependence of this ferromagnetic remnant moment $m_R$(T) does not show any change in slope near 9 K. Note also that the slope of the ferromagnetic $m_R(T)$ curve is approximately the same as of the other three curves for $T > T_c$ and, that the remnant moment depends on the value of the FC applied. The behavior of the three $m_R(T)$ curves of figure 5(b) can be understood if the superconducting state exists simultaneously with a ferromagnetic background state. In figure 6(a) we show some of the results of hysteresis loop measurements at 5 K for H // c between the fields ± 500 Oe and ± 1000 Oe. For loops between higher fields the signal becomes significantly noisier in the field interval 1000 < H ≤ 4500 Oe, the origin of which is probably due to flux creep in the superconducting magnet that causes a voltage drift in the SQUID magnetometer (Quantum Design® 2000). Independently of this, an up turn and down turn of the hysteresis loop extremities as the field is increased is clearly evident (see figure 6(a)). For fields in the vicinity of 10 000 Oe, the distortion of the hysteresis loops are so large that a ferromagnetic-like background is revealed under the superconducting loops (see inset of figure 6(a)). These hysteresis loops were obtained after the subtraction of a linear diamagnetic background $\chi(5\ K) = -7.52 \times 10^{-6}$ emu $g^{-1}$ $Oe^{-1}$ for ± 10 000 Oe and $\chi(5\ K) = -7.02 \times 10^{-6}$ emu $g^{-1}$ $Oe^{-1}$ for ± 500 Oe and ± 1000 Oe.

After five and a half months of constant and reproducible results obtained from each of the two pieces of the sample, a large change in the magnetic response occurred spontaneously in both pieces at the same time. No contact or manipulation was made with the samples since they were mounted in their sample holders (plastic straws) and stored in an argon atmosphere most of the time. Figure 6(b) shows the hysteresis loops before (inset figure 6(a)) and after this change (the diamagnetic background becomes $\chi(5\ K) = -6.9$ emu $g^{-1}$ $Oe^{-1}$). Note that the ferromagnetic character of the sample is significantly enhanced. A large hysteresis and remnant moment are clearly evident and the saturation magnetic moment rose from $m_s \approx 0.0013$ emu/g for H ≥ 3800 Oe to $m_s \approx$ 0.019 emu/g for H ≥ 13 000 Oe (measurements were also performed up to 70 000 Oe). The upper inset of figure 6(b) shows the direct hysteresis loops m(H), measured at 5 K and 12 K (H // c) after the magnetic response of the sample changed, for fields between ± 500 Oe. This figure shows that at T = 5 K, in addition to the large ferromagnetic loop that is slightly distorted by the diamagnetic background, the sample still exhibits superconductivity as revealed by the initial data points of the loop where a virgin magnetization curve with a Meissner line is clearly evident. This superconducting state cannot be seen in the hysteresis loop at T = 12 K since the sample is then in the normal state. Interestingly, but not unexpected, subtracting the 12 K hysteresis loop from the 5 K loop gives exactly the same superconducting hysteresis loop shown in figure 2. The lower inset of figure 6(b) presents the ZFC and FCC temperature dependencies of the magnetic moment m(T) for H = 50 Oe (H // c), after the magnetic character of sample had changed. Note the paramagnetic background in the normal state of this sample. This figure confirms that the sample remains superconducting with approximately no change in the $T_c$ onset or in the superconducting volume fraction. In other words, the spontaneous change in the sample involves an increase in the ferromagnetic saturation moment by approximately fifteen times, the character and volume fraction of the superconducting state remains nearly constant.

The magnetic properties of the graphite-sulfur samples presented herein (figures 1-3) are completely consistent with the occurrence of superconductivity below 9 K, although no 'zero resistance' state could be established by means of transport measurements. The failure to observe zero resistance may be due to several factors: the small superconducting volume fraction; the occurrence of induced superconductivity by proximity effect that suggests the existence of localized superconducting islands or grains that may not form a percolating path; the weak coupling of the pressed powder sample or the presence of insulating sulfur inclusions as indicated by the larger (~ 2000x) resistance of the sample; or a combination of these effects. Besides the occurrence of superconductivity in these samples, another important result is the large anisotropy of the superconducting state. This is revealed by the results of figure 4(a) where no superconducting signal was detected (within the data noise of ~ 5 x $10^{-6}$ emu/g) when the magnetic field was applied parallel to the graphite planes (H // a). This means that the superconducting diamagnetic moment is at least 10x smaller for H // a than for H // c. The observed anisotropy strongly suggests that the superconductivity is associated with the graphite planes. This could also explain the difficulty in detecting the superconducting signal for the H // a configuration since the graphite planes transverse contribution is expected to be small. The ferromagnetism is present in both configurations H // a and H // c for T < $T_c$, as shown in figures 4(b) and 6. A ferromagnetic state was also observed for temperatures T > $T_c$ up to room temperature. Ferromagnetic behavior was even found in our initial bulk graphite material but with a weaker saturation magnetic moment ($m_s \approx$ 1.8 x $10^{-4}$ emu/g). Since impurities are usually suspected as being responsible for the observed ferromagnetism, special care was taken select starting materials of the highest purity available and to avoid possible contamination during samples preparation and measurement. Besides, we have already shown (Kopelevich et al. 2000) that the initial magnetic impurities are not responsible for the ferromagnetic response of pure graphite. Our graphite has less than 5 ppm of impurities; if we assume that all impurities are magnetic and concentrated in a cluster, we calculate a saturation magnetic moment of ~ 6.8 x $10^{-4}$ emu/g. This value is about twice as large as the magnetic moment found in our graphite-sulfur samples, but also 28 times smaller than the saturation moment measured after the sample changed spontaneously. Even though the existence of a magnetic cluster could explain the initial magnetic response of our sample, a very unlikely transformation like a spontaneous reduction of a ferromagnetic impurity such as $Fe_3O_4$, would have to occur to account for what happened in our final samples. Similarly, given the small amount of impurities of our samples, it is also difficult to account for the superconductivity and its anisotropic nature in terms of impurities. In other words, the ferromagnetism and the superconductivity appear to be intrinsic properties of samples. It is noteworthy that a number of theoretical papers (Nakada et al. 1996, Harigaya 2001, Khveshchenko 2001, Gonzalez et al. 2001, Singh 2002) have appeared in which the occurrence of magnetic and superconducting correlations in graphite have been predicted. We believe that at this stage of the experimental research, it is too premature to speculate about the mechanisms responsible for the spectacular phenomena shown here. However, we would like to present a possible scenario for the simultaneous occurrence of ferromagnetism and superconductivity and make some comments about the main results.

The observed simultaneous occurrence of diamagnetism, ferromagnetism and superconductivity (for $T < T_c$) in an apparently independent form may not be so surprising if we take into account that all contributions to the magnetic moment of graphite can come from different charged carriers near the Fermi level (Sharma et al. 1974). The ferromagnetism seems to be intrinsic to pure graphite and occurs at all temperatures smaller than the Curie temperature (Kopelevich et al. 2000, Moehlecke et al. 2002). In the graphite-sulfur samples, the ferromagnetism also occurs and coexists with superconductivity for $T < T_c$ which seems to be localized in regions of the graphite planes stabilized by sulfur. We speculate that the spontaneous increase of ferromagnetism could be related to a rearrangement of the adsorbed sulfur to optimal positions or to sulfur desorption. Clearly, further studies are necessary to verify these hypotheses.

## Acknowledgements

This work was supported by FAPESP, CNPq and CAPES Brazilian science agencies and the US Department of Energy under Grant No. DE-FG03-86ER-95230.

## Figure captions

Figure 1. (a) Temperature dependencies of the magnetic moment measured in zero-field-cooled, $m_{ZFC}(T)$, and field-cooled on cooling, $m_{FCC}(T)$, in two applied magnetic fields, 10 and 70 Oe, for $H // c$. (b) Normalized ZFC magnetic moment measured at various fields: (▼) - $H = 30$ Oe; (◊) - $H = 100$ Oe; (▲) - $H = 300$ Oe; (○) - $H = 500$ Oe; (■) - $H = 2000$ Oe. Arrows indicate the superconducting transition temperature $T_c(H)$ onset.

Figure 2. ZFC magnetic moment hysteresis loops $m(H)$ for $T = 2$ K (◊), $T = 5$ K (□) and $T = 6$ K (○), with $H // c$, after the subtraction of their diamagnetic backgrounds. For details, see text. The inset presents the direct $m(H)$ data measured for $T = 2$ K ($H // c$).

Figure 3. (a) Initial low field portion of hysteresis loops $m(H)$ for temperatures between 2.0 K and 7.5 K (ZFC, $H // c$). (b) Temperature dependence of the lower critical field,

$H_{c1}(T)$, and the breakdown field, $H_b(T)$. The inset shows the calculated penetration depth, $\lambda(T)$, temperature dependence. See text for details about the fitted lines.

Figure 4. (a) ZFC (□) and FCC (●) temperature dependencies of the magnetic moment m(T) for H // a and different magnetic fields, as indicated in Oe next to each curve. (b) ZFC hysteresis loops for T = 2 K (■) and 9 K (○) after the subtraction of the diamagnetic background (H // a). The upper inset presents the raw data for the measured hysteresis loop at T = 2 K (H // a). The lower inset shows the hysteresis loop at 5 K up to high magnetic fields (H // a).

Figure 5. (a) Temperature dependence of the remnant moment, $m_R(T)$, after the sample was field cooled under 1000 Oe (H // a). (b) Temperature dependencies of the remnant moment, $m_R(T)$, after the sample was field cooled under the fields (in Oe) indicated next to each curve (H // c). Also shown in this figure are the initial $m_R(T)$ data for H // a of figure 5(a).

Figure 6. (a) Hysteresis loops measured at 5 K (H // c) for fields between ± 500 Oe and ± 1000 Oe, after subtraction of the diamagnetic background. The inset shows the same kind of loop (5 K, H // c) between ± 10 000 Oe. (b) Hysteresis loops measured at 5 K (H // c) between ± 10 000 Oe before (◇) and after (■) the magnetic properties of the sample changed and with the backgrounds subtracted. The upper inset shows the direct hysteresis loops measured at 5 K and 12 K (H // c) between ± 500 Oe, after the magnetic properties

of the sample changed. The lower inset presents the ZFC and FCC magnetic moment measured for H = 50 Oe (H // c), after the magnetic properties of the sample changed.

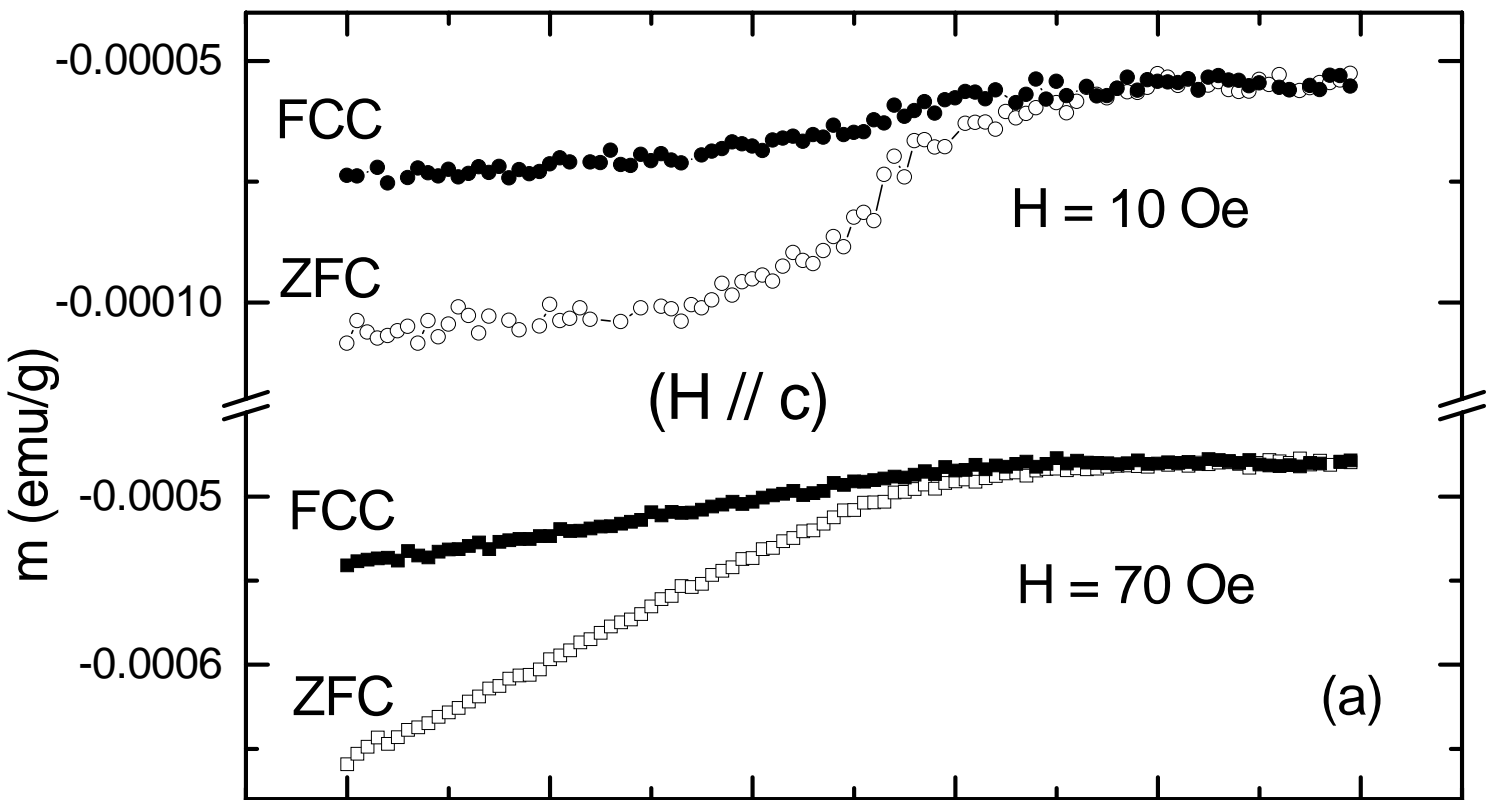
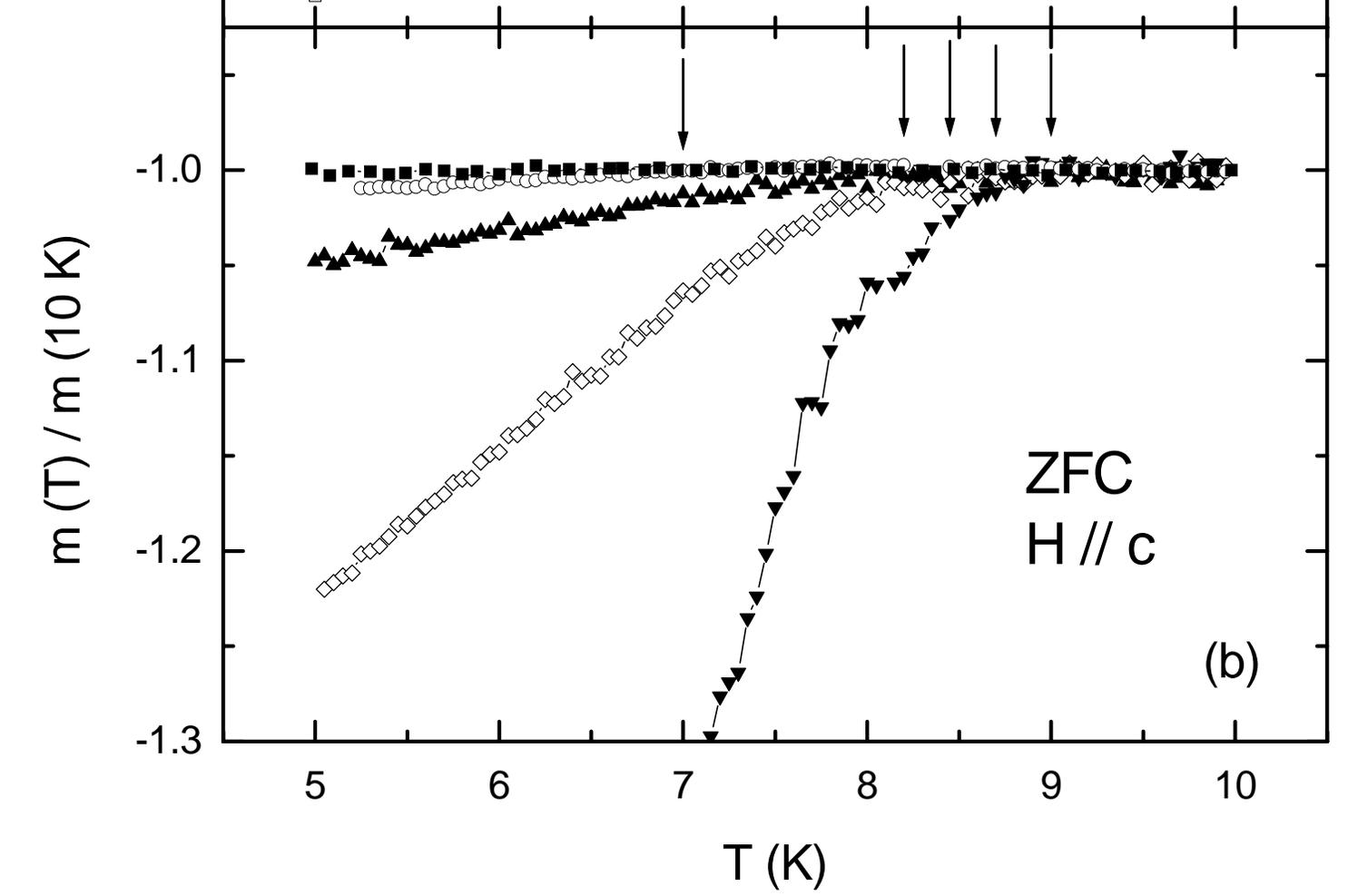

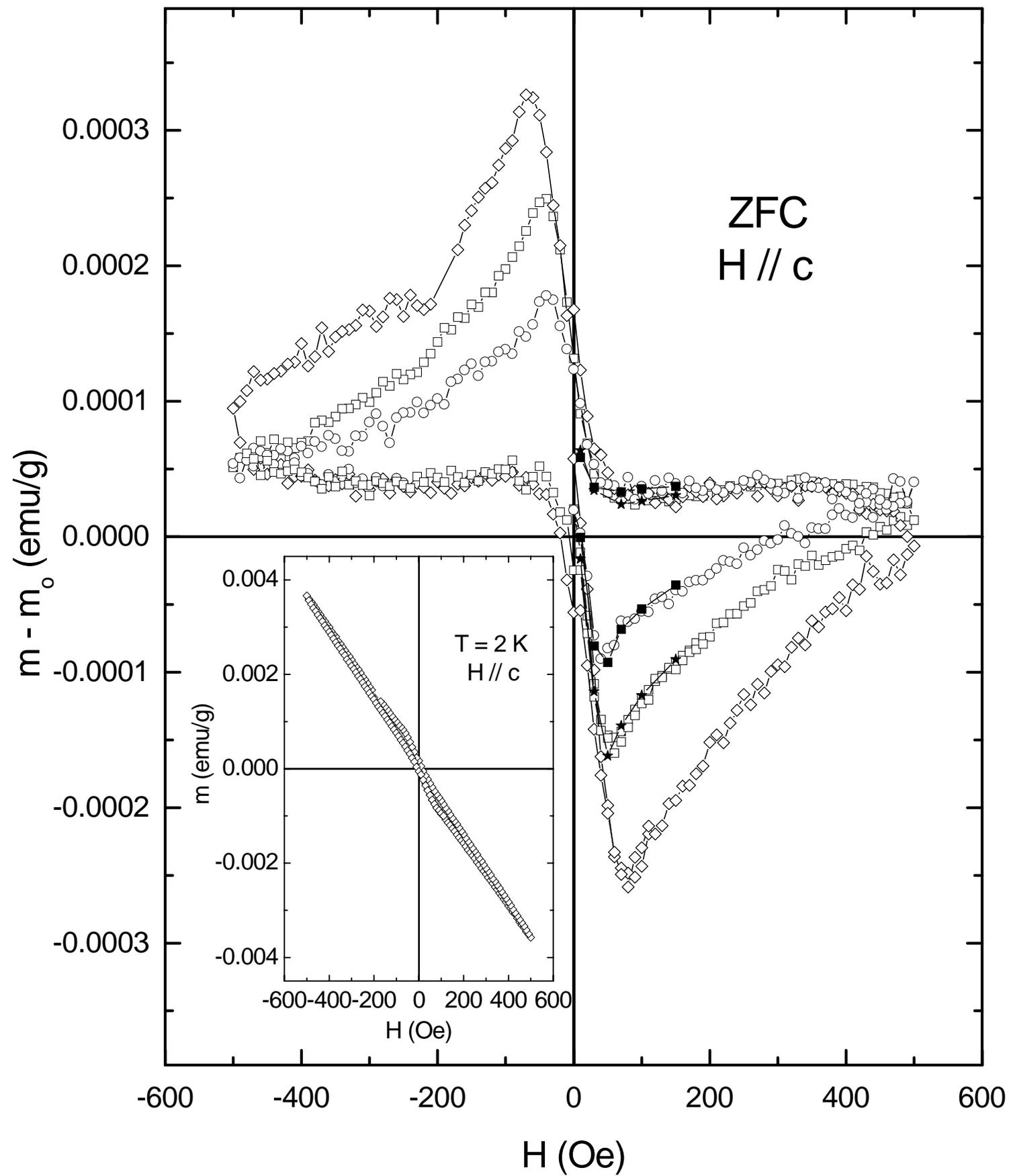

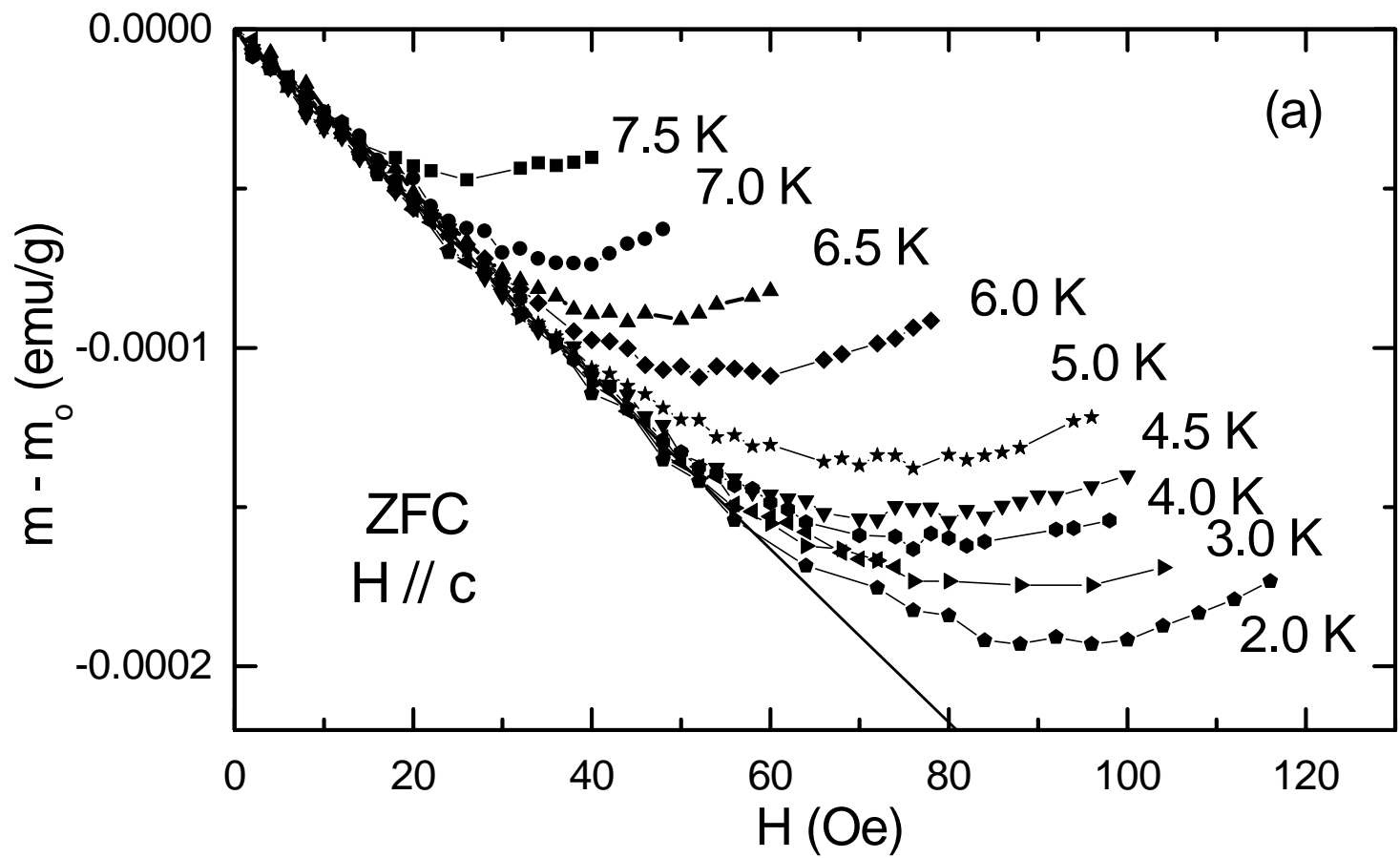
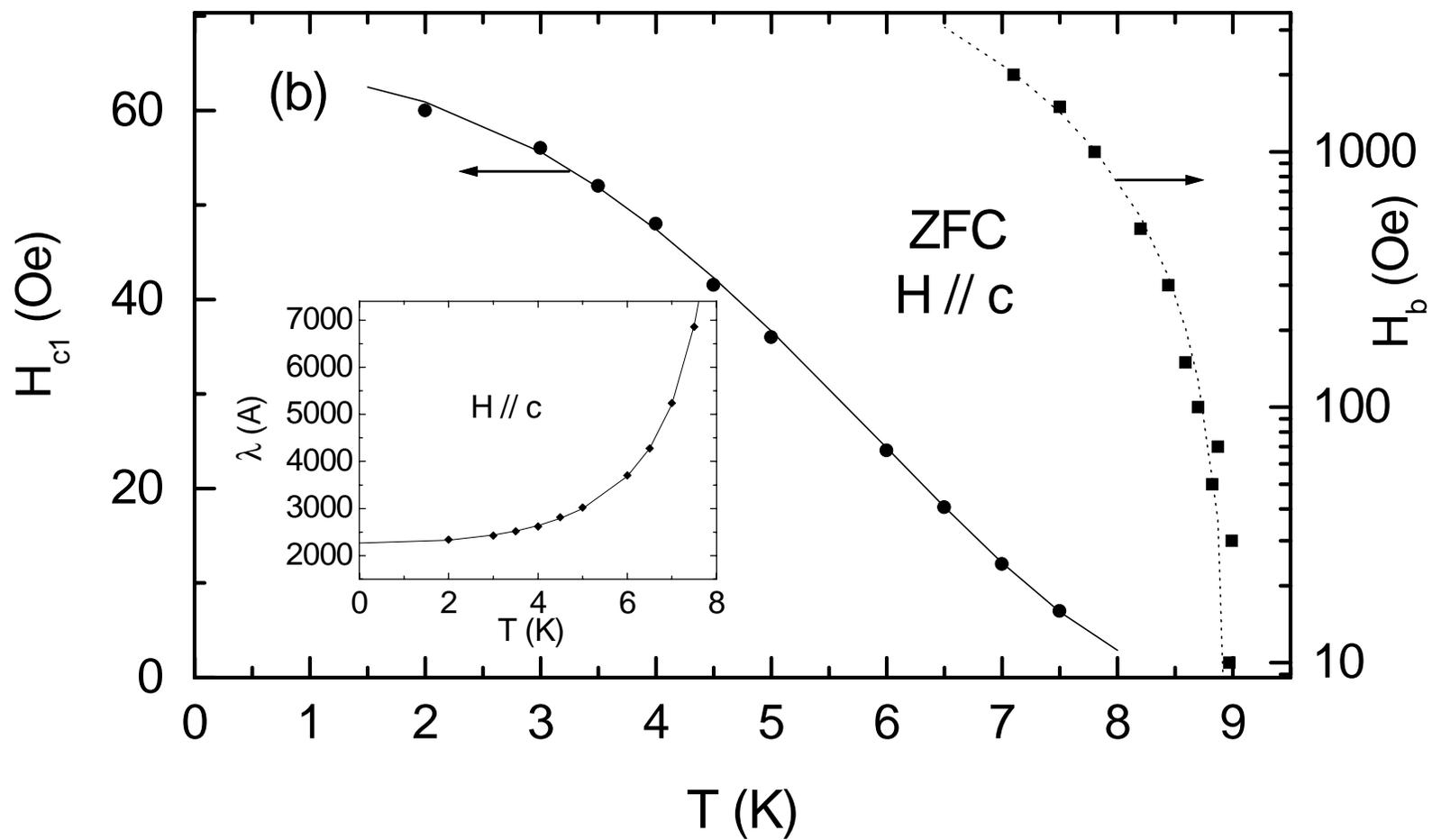

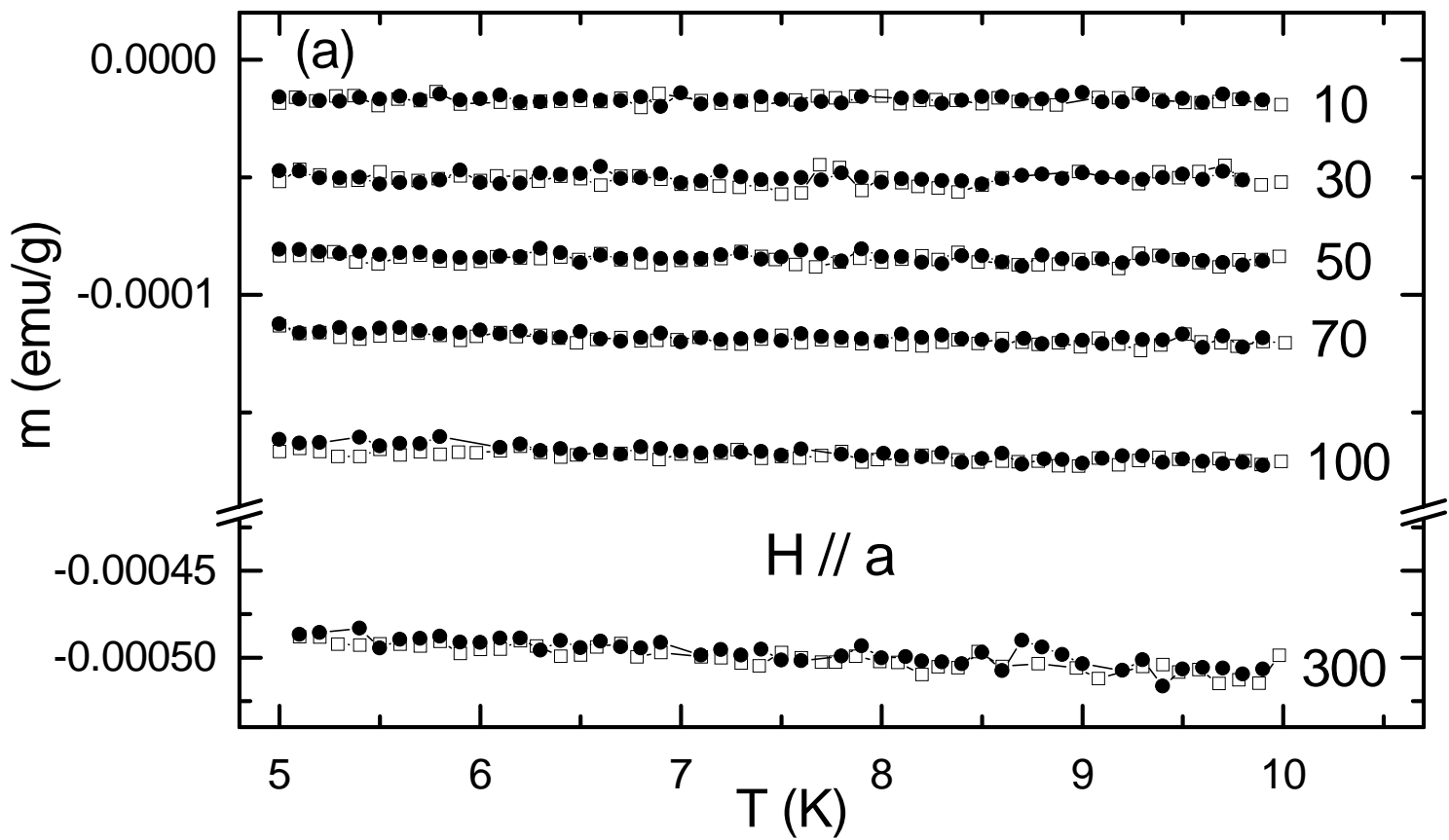
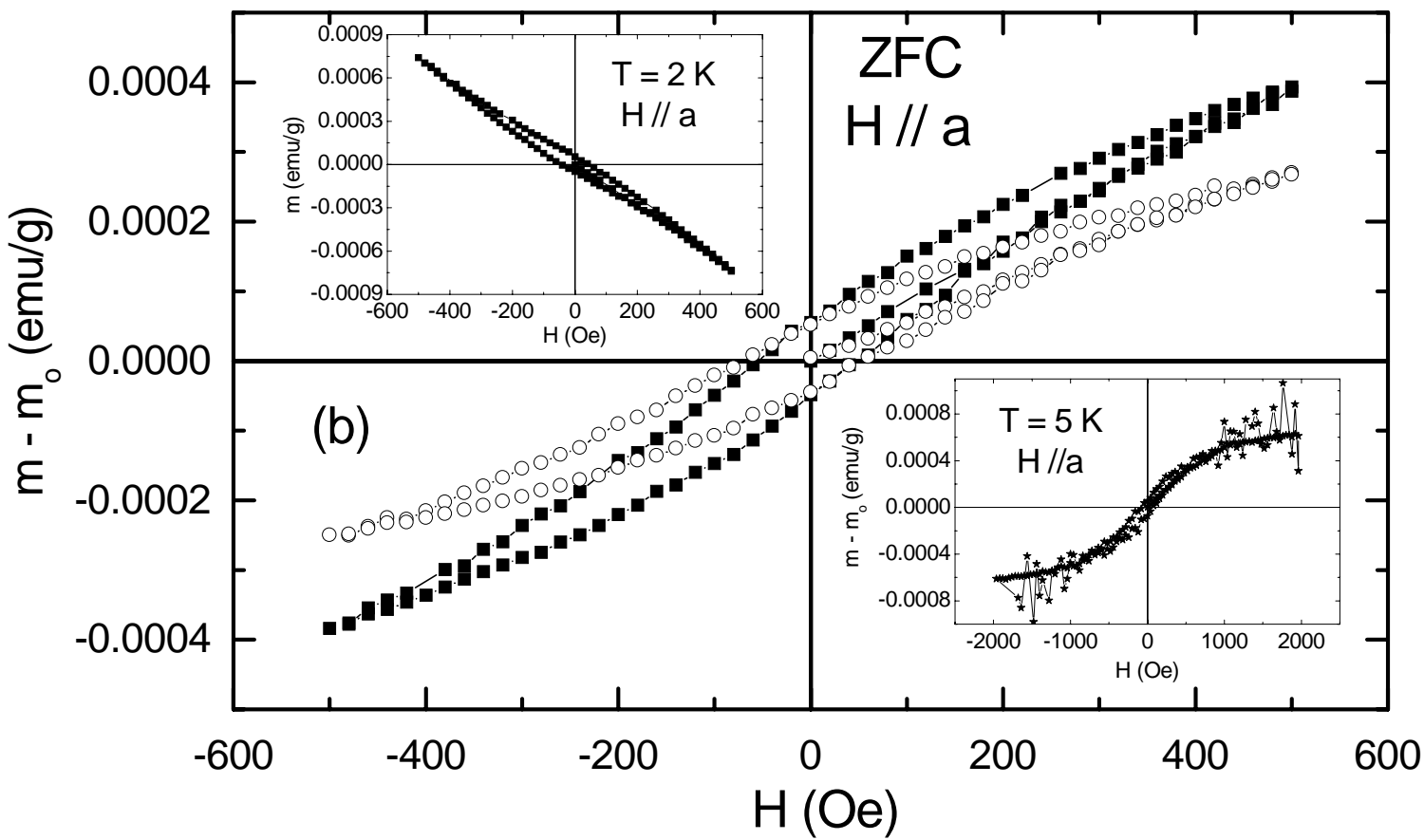

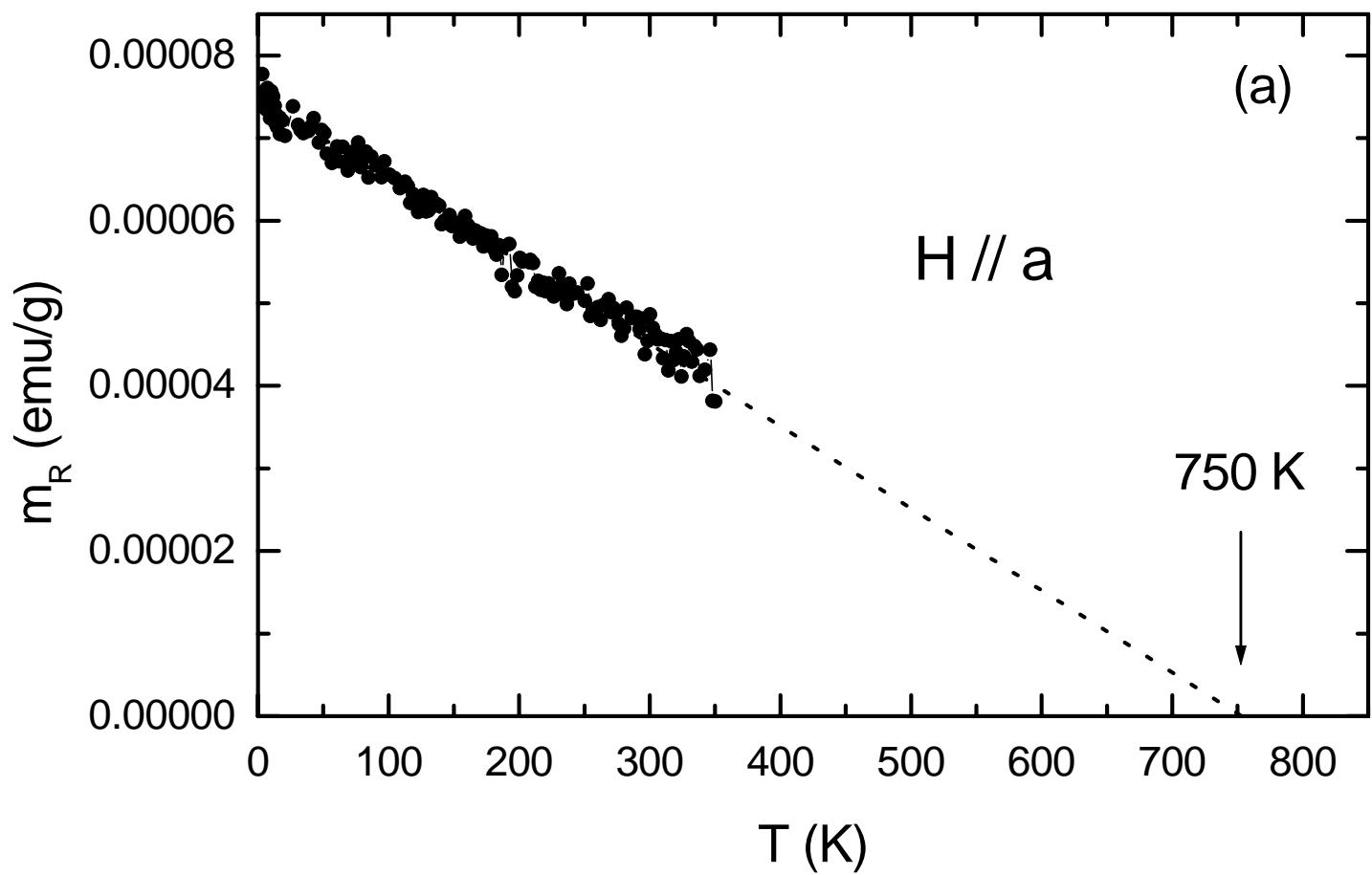
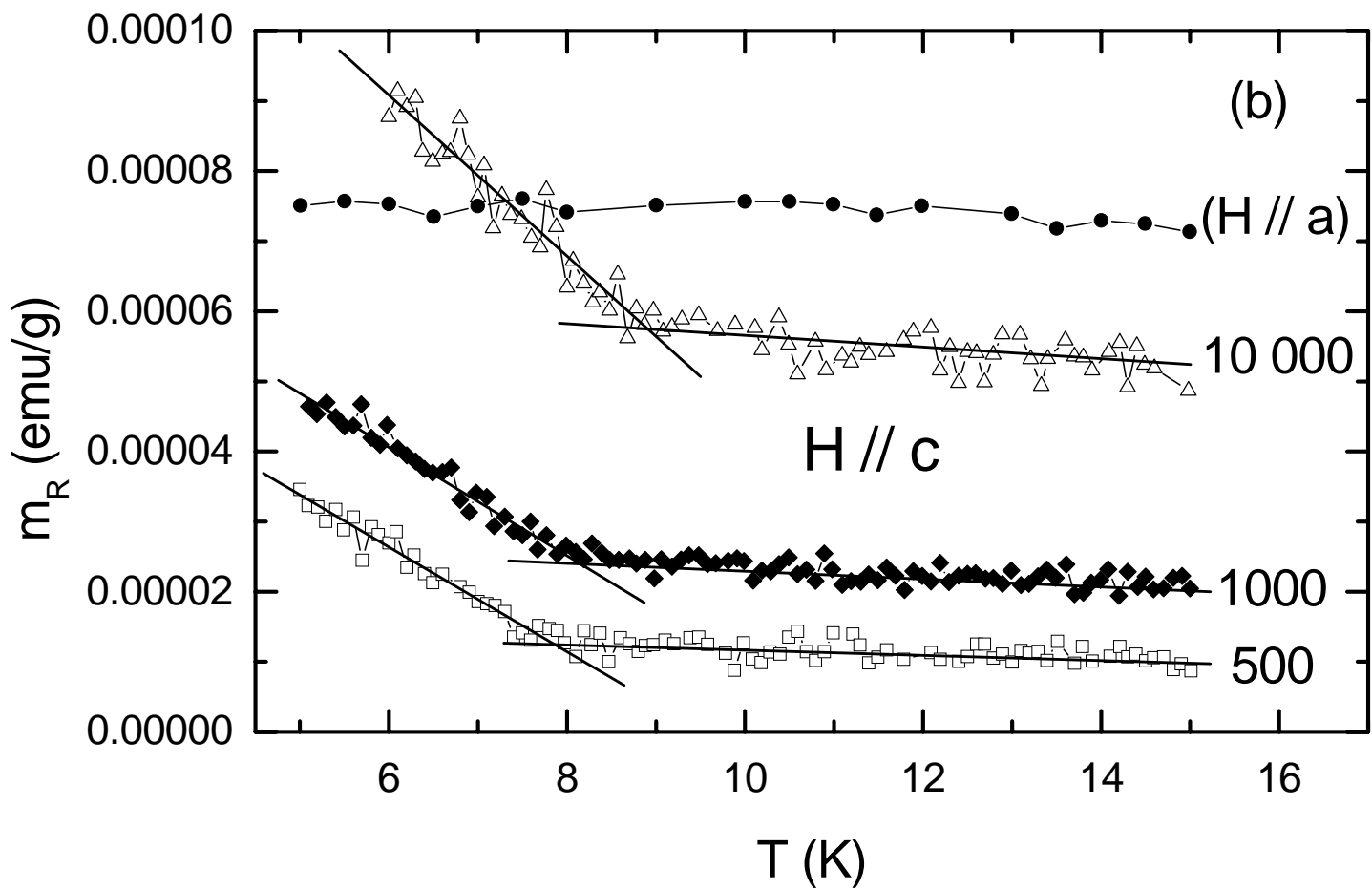

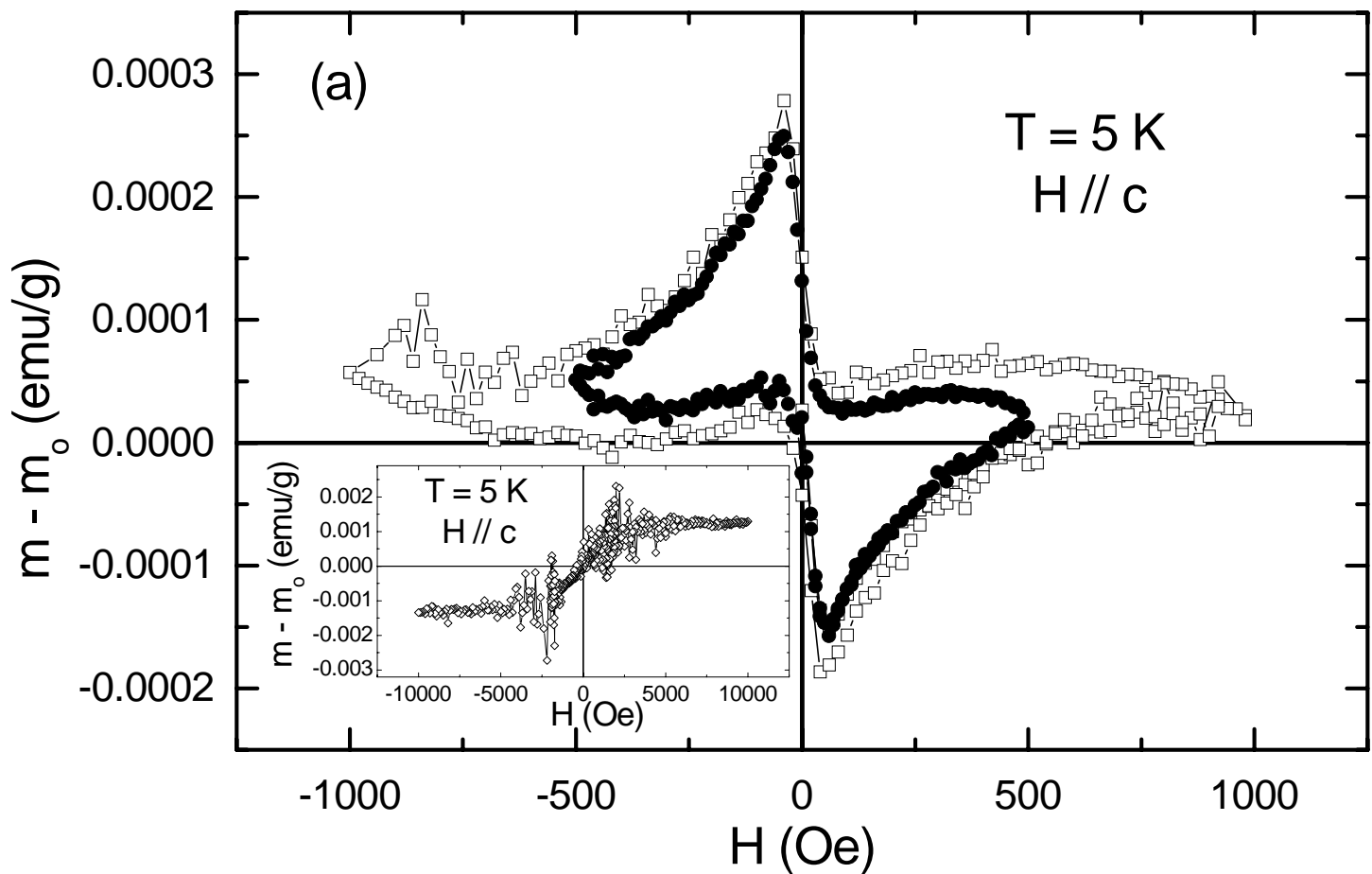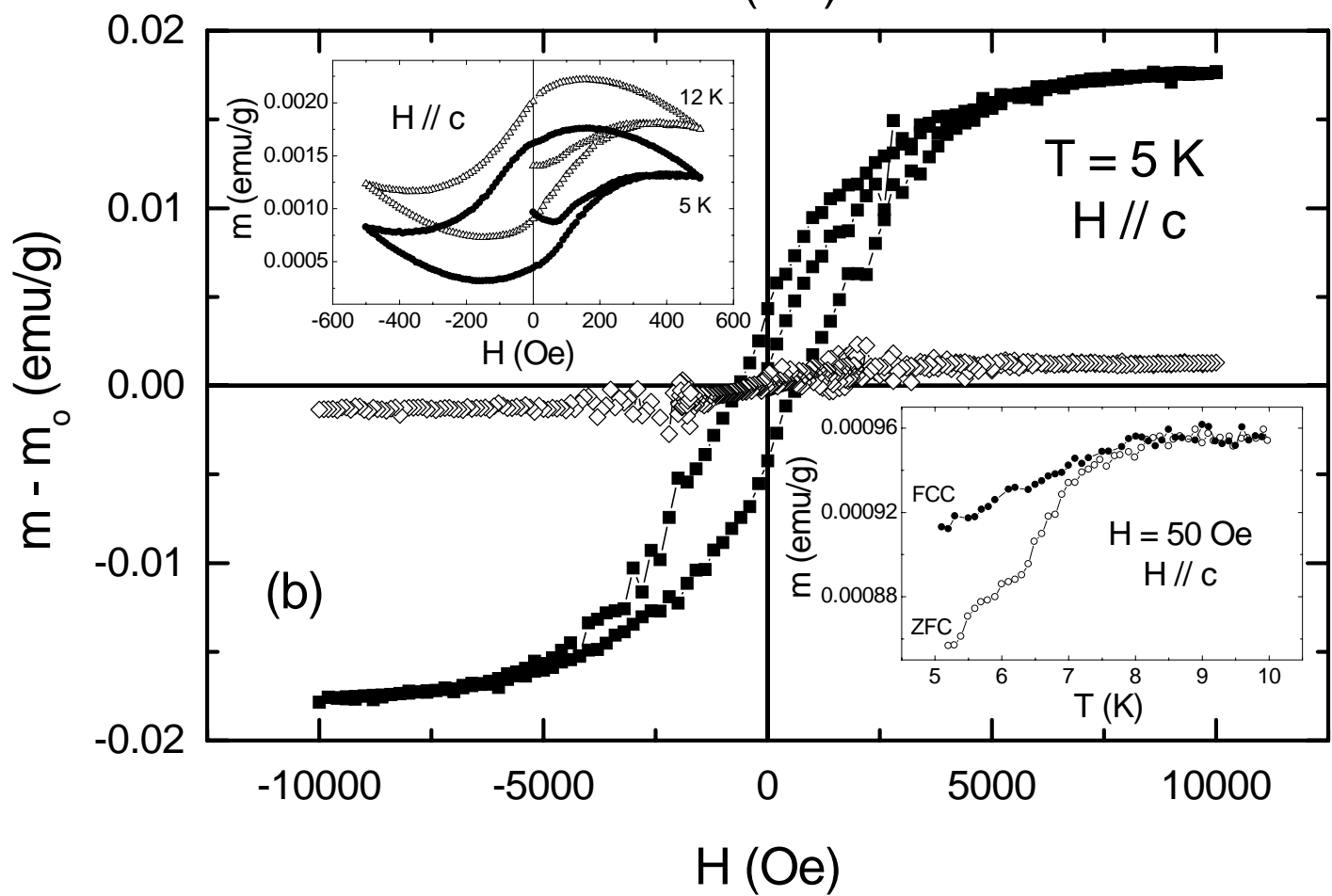